\documentclass{ws-ijmpe} %For IJMPE
\usepackage[super,compress]{cite}
\begin{document}

\markboth{V. V. Pilipenko, V. I. Kuprikov}{Model of multiple Dirac eikonal scattering of protons by nuclei}

%--change \catchline{}{}{}{}{}
%%%%%%%%%%%%%%%%%%%%%%%%%%%%%%%%%%%%%%%%%%%%%%%%%%%%%%%%%%%%%%%%%%%%

\title{Model of multiple Dirac eikonal scattering of protons by nuclei}

\author{V. V. Pilipenko and V. I. Kuprikov}

\address{National Science Center ``Kharkov Institute of Physics and Technology'',
\\ Kharkov 61108, Ukraine \\ vpilipenko@kipt.kharkov.ua}

%\author{Second Author}

%\address{Group, Laboratory, Address\\
%City, State ZIP/Zone, Country\\
%second\_author@group.com}

\maketitle

%%\begin{history}
%%\received{Day Month Year}
%%\revised{Day Month Year}
%\accepted{Day Month Year}
%\comby{(xxxxxxxxxx)}
%%\end{history}

\begin{abstract}
The model of multiple Dirac eikonal scattering of incident proton by target-nucleus nucleons is developed, in which new expressions for the elastic $pA$-scattering amplitudes are obtained from the multiple scattering Watson series with employing the eikonal approximation for the Dirac propagators of the free proton motion between successive scattering acts on nucleons. Basing on this model, calculations for the complete set of observables of the elastic $p+^{40}$Ca and $p+^{208}$Pb at 800 MeV have been performed with using proton-nucleon amplitudes determined from the phase analysis and the nucleon densities obtained from describing the target-nucleus structure in the relativistic mean-field approximation. A comparison has been made of the results of these calculations with analogous calculations on the basis of the Glauber multiple diffraction theory.
\end{abstract}


\ccode{PACS numbers: {24.10.-i,25.40.Cm,11.80.La}}

%\tableofcontents

\section{Introduction}

The multiple diffraction scattering theory (MDST), the groundwork of which was laid by the initial paper~\cite{Gla1} of Glauber is a fairly successful and popular approach to describing the processes of scattering of protons and other hadrons on atomic nuclei in the intermediate energy region. This approach has found its development and employment in a great number of works of many authors and, in particular, different refinements and corrections to the initial MDST formulation were considered and discussed (see, for example, Refs.~\citen{Gla2,Alk,Sat,Sit,Osl,Wall1,Wall2,Ble1,Ble2,Kup1,Pil1,Kup2,Ber} and references therein). At the present time, another approach also often employed for describing the intermediate-energy proton scattering by nuclei is the model of relativistic impulse approximation, which is based on solving the Dirac equation with a relativistic microscopic optical potential (see, for example, Refs.~\citen{McN,Ray,Clar} and references therein). Although the MDST does not lose its applicability and is succesfully employed to analyzing various scattering processes of protons and nuclei, it seems interesting to consider a consistent relativistic development of this approach basing also on the Dirac equation with simultaneously employing a relativistic description of the structure of target nuclei.

At present, in the literature a number of precise measurements are available for differential cross sections and spin observables of the proton-nucleus ($p$--$A$) scattering at intermediate energies (namely, the order of several hundred MeV). Analysis of these experimental data can be used for obtaining valuable information about the structure of nuclei. However, for performing a reliable analysis of these data sufficiently accurate theoretical methods are needed. In Refs.~\citen{Kup1,Pil1,Kup2,Ber}, we obtained and tried out some rather sophisticated expressions for the $p$--$A$ scattering amplitudes in the framework of MDST basing on the realistic $NN$-amplitudes and nuclear densities and allowing for two-nucleon correlations by means of including intermediate excitations of nuclei as well as taking account of noneikonal corrections. Notwithstanding the certain success of this consideration, some shortcomings in the description of the analyzed spin observables may indicate that possible improvements of the model should be sought for.

In this article, we carry out the development of the MDST approach by means of constructing a relativistic model of the multiple scattering of the incident proton on the nucleons of the target nucleus basing on the eikonal approximation for the Dirac equation, which we will call the multiple Dirac eikonal scattering model (MDES). The new expressions for the $p$--$A$ scattering amplitude obtained by us in this approach are employed for calculations of the complete set of observables for the elastic $p$--$A$ scattering with using the realistic nucleon densities calculated on the basis of approximation of relativistic mean field (RMF) (see, for example, Refs.~\citen{Wale,Hor,Gam,Ser,Meng,Typ,Lal}) in comparison with the analogous calculations by the MDST.

\section{Description of the MDES model}

We may suppose that for further improving this approach it would be probably advisable to employ a more consistent relativistic description of the process of multiple scattering of the incident proton on nucleons of the target nucleus, which we are going to develop basing on the eikonal approximation for the Dirac equation. This is analogous to considering MDST as the theory of the multiple eikonal scattering of the incident particle on nucleons in the nucleus. Therefore, we will build our model in the way close to the approach of Refs.~\citen{Ble1,Ble2}, which was formulated taking account of the results of the works of Refs.~\citen{Wall1,Wall2}. This approach is based on a number of assumptions, which allow for the mutual cancelation of different corrections to the MDST, namely: one may neglect the motion of nucleons in the nucleus during the interaction with the incident proton, as well as the contribution of rescatterings of the projectile on the same nucleon; the most essential matrix elements of the $t$-operator may be considered to be local and determined on the energy shell. Note that in Ref.~\citen{Ama} the Dirac-eikonal amplitude was considered for the proton-nucleus scattering described by a phenomenological potential

We will proceed from the representation of the $p$--$A$ scattering $T$-matrix operator in the form of the multiple scattering Watson series,\cite{Gold} neglecting here rescatterings on the same target nucleons and therefore restricting the series up to the terms of order $A$:
%\begin{equation}
\begin{eqnarray}
T &=& \sum\limits_{j = 1}^A {t_j }  + \sum\limits_{j = 1}^A {\sum\limits_{k \ne j}^A {t_j \tilde G^{( + )} t_k } } \nonumber \\
&& + \sum\limits_{j = 1}^A {\sum\limits_{k \ne j}^A {\sum\limits_{l \ne k \ne j}^A {t_j \tilde G^{( + )} t_k \tilde G^{( + )} t_l } }  + ... \mbox{+ term\;of\;order\;{\it A}} },
\label{Eq1}
\end{eqnarray}
%\end{equation}
where $t_j $ are the $t$-matrix operators for the proton scattering on individual nucleons of the target nucleus, which are considered to be the same as for scattering on free nucleons in the impulse approximation spirit. We assume that the quantities in Eq.~(\ref{Eq1}) are written in the $p$--$A$ center-of-mass (c.m.) frame. The propagator of free motion of the proton between the scattering acts on nucleons is taken in the form
\begin{equation}
\tilde G^{( + )}  = \left[ {E - H_p \left( {{\bf{\hat k}}} \right) - H_t \left( {{\bf{\hat k}},\left\{ {{\bf{r}}_j } \right\}} \right) + i0} \right]^{ - 1},
\label{Eq2}
\end{equation}
\begin{equation}
H_p \left( {{\bf{\hat k}}} \right) = {\bm{\alpha}\bf{\hat k}} + \beta m , \; H_t \left( {{\bf{\hat k}},\left\{ {x_j } \right\}} \right) = \sqrt {{\bf{\hat k}}^2  + h_t ^2 \left( {\left\{ {x_j } \right\}} \right)} .
\label{Eq3}
\end{equation}
Here, ${\bf{\hat k}} = {\bf{\hat k}}_p  =  - {\bf{\hat k}}_t  =  - i{\partial / \partial \bf{r}}$ is the momentum operator, ${\bf{r}} = {\bf{r}}_p  - {\bf{r}}_t $, ${\bf{r}}_p $ and ${\bf{r}}_t $ are the coordinates of– the proton and the nucleus as a whole (we assume $\hbar  = c = 1$). The Hamiltonian of the incident proton $H_p $ has the usual Dirac form, while the Hamiltonian $H_t $ of the target in the case of scattering on a spinless nucleus is presented in a model form,~\cite{Rem} in which $h_t \left( {\left\{ {x_j } \right\}} \right)$ is the Hamiltonian of the internal motion of the nucleus ($\left\{ {x_j } \right\}$ is the set of internal nucleon variables), which has to provide relativistic kinematic relations when taking account of the target recoil. The system energy in the $p$--$A$ c.m. frame is $E = \varepsilon _p  + \varepsilon _t $, where the energies of proton and nucleus are $\varepsilon _p  = \sqrt {k^2  + m^2 } $ and $\varepsilon _t  = \sqrt {k^2  + M^2 } $, $m$ and $M$ being their masses; $k = \left| {{\bf{k}}_i } \right| = \left| {{\bf{k}}_f } \right|$ is the magnitude of the initial and final momenta.

The amplitude of the elastic $p$--$A$ scattering is related to the matrix element of the operator~(\ref{Eq1}) in the $p$--$A$ c.m. frame by the following formula:
\begin{equation}
F\left( {{\bf{k}}_f ,{\bf{k}}_i } \right) =  - \frac{1}{{4\pi }}\frac{{\varepsilon _t }}{{\varepsilon _p  + \varepsilon _t }}\left( {u^{\mu '} \left( {{\bf{k}}_f } \right)\Psi _0^ +  \left\langle {{\bf{k}}_f } \right|T\left| {{\bf{k}}_i } \right\rangle \Psi _0 u^\mu  \left( {{\bf{k}}_i } \right)} \right) .
\label{Eq4}
\end{equation}
Here, $\Psi _0 $ is the wave function of the ground state of the nucleus; $\left| {{\bf{k}}_{i,f} } \right\rangle $ are the vectors of the initial and final plane waves normalized to the unit particle density; the proton bispinors $u^\mu  \left( {\bf{k}} \right)$ are normalized according to the formula: $\bar u^\mu  \left( {\bf{k}} \right) \cdot u^\mu  \left( {\bf{k}} \right) = 2m$.

In Eq.~(\ref{Eq1}) the $j$-th nucleon in the nucleus is displaced by the vector ${\bf{r}}_j $ relative to the coordinate ${\bf{r}}_t $ of the nucleus as a whole. Therefore, if $t^{(j)} $ is the $t$-operator for the scattering on a nucleon of the sort $j$ placed in the origin of coordinates, then for the $t$-matrix element for the scattering on the $j$-th nucleon in the nucleus we have: $\left\langle {{\bf{k'}}} \right|t_j \left( {{\bf{r}}_j } \right)\left| {\bf{k}} \right\rangle  = \exp \left( {i{\bf{qr}}_j } \right)\left\langle {{\bf{k'}}} \right|t^{(j)} \left| {\bf{k}} \right\rangle $, where ${\bf{q}} = {\bf{k}} - {\bf{k'}}$. It is necessary to relate the matrix element $\left\langle {{\bf{k'}}} \right|t^{(j)} \left| {\bf{k}} \right\rangle $ with the amplitude of the $p$--$N$  scattering determined in the $p$--$N$  c.m. frame. As it was noted above we may assume that the main role is played by the matrix element values of $t^{(j)} $, determined on the energy shell. In this case, we may consider these matrix elements to be local, i.e. such that depend only on the momentum transfer ${\bf{q}}$ and the fixed energy corresponding to the energy shell (see Refs.~\citen{Ble1,Ble2}). For the off-energy-shell the matrix element values it will be acceptable to use the same approximation, as that determined in the $p$--$N$  c.m. frame on the energy shell.

In the noncovariant form, the $pN$-amplitude $f_j \left( {{\bf{q}}_c } \right)$ (here, ${\bf{q}}_c  = {\bf{k}}_c  - {\bf{k'}}_c $ is the momentum transfer in the $p$--$N$ c.m. frame) has the following form (see, for example, Ref.~\citen{McN}):
%\begin{equation}
\begin{eqnarray}
f_j \left( {{\bf{q}}_c } \right) &=& A_j \left( {q_c } \right) + B_j \left( {q_c } \right){\bm{\sigma }}_p {\bm{\sigma }}_j  + C_j \left( {q_c } \right)\left( {{\bm{\sigma }}_p  + {\bm{\sigma }}_j } \right){\bf{n}} \nonumber \\
&& + D_j \left( {q_c } \right)\left( {{\bm{\sigma }}_p {\bf{q}}_c } \right)\left( {{\bm{\sigma }}_j {\bf{q}}_c } \right) + E_j \left( {q_c } \right)\sigma _{pz} \sigma _{jz} .
\label{Eq5}
\end{eqnarray}
%\end{equation}
Here, ${\bm{\sigma }}_p $ and ${\bm{\sigma }}_j $ are the Pauli matrices of the incident proton and of the nucleon in the nucleus; the direction ${\bf{e}}_z $ is chosen along the average momentum $
{\bf{k}}_{ca}  = ({\bf{k}}_c  + {\bf{k'}}_c )/2$, and the normal to the scattering plane is ${\bf{n}}_c  = {\bf{k}}_c  \times {\bf{k'}}_c /|{\bf{k}}_c  \times {\bf{k'}}_c | = {\bf{\hat q}}_c  \times {\bf{e}}_z $, where ${\bf{\hat q}}_c  = {\bf{q}}_c /q_c $, ${\bf{q}}_c  \bot {\bf{k}}_{ca} $. The amplitude~(\ref{Eq5}) is a spin operator, whose matrix elements are calculated between the spinors of proton $\chi _p $ and nucleon $\chi _j $, which are normalized to unity. In the noncovariant form the matrix element of $t^{(j)} $ is related to the amplitude $f_j \left( {{\bf{q}}_c } \right)$ by the formula:
\begin{equation}
\left\langle {{\bf{k'}}_c } \right|t^{(j)} \left| {{\bf{k}}_c } \right\rangle _{n{\mathop{\rm c}\nolimits} }  =  - \frac{{2\pi }}{{\varepsilon _{pc} \varepsilon _{Nc} }}\left( {\varepsilon _{pc}  + \varepsilon _{Nc} } \right)f_j \left( {{\bf{q}}_c } \right) =  - \frac{{4\pi }}{{\varepsilon _{pc} }}f_j \left( {{\bf{q}}_c } \right) .
\label{Eq6}
\end{equation}

Let us introduce the invariant Dirac amplitude of the $p$--$N$ scattering in the form (see, for example, Ref.~\citen{McN}):
%\begin{equation}
\begin{eqnarray}
\hat F\left( {s,q_c^2 } \right) &=& F_S \left( {s,q_c^2 } \right) + F_V \left( {s,q_c^2 } \right)\gamma _p^\mu  \gamma _{j\mu }  + F_T \left( {s,q_c^2 } \right)\sigma _p^{\mu \nu } \sigma _{j\mu \nu } \nonumber \\
&& + F_P \left( {s,q_c^2 } \right)\gamma _p^5 \gamma _j^5  + F_A \left( {s,q_c^2 } \right)\gamma _p^5 \gamma _p^\mu  \gamma _j^5 \gamma _{j\mu } .
\label{Eq7}
\end{eqnarray}
%\end{equation}
Here, $\gamma _p^\mu  $ and $\gamma _j^\mu $ are the Dirac matrices of the proton and nucleon, $\gamma ^5  = i\gamma ^0 \gamma ^1 \gamma ^2 \gamma ^3 $; $\sigma ^{\mu \nu }  = i\left[ {\gamma ^\mu  ,\gamma ^\nu  } \right]/2$ are the Dirac tensor operators; $F_S $, $F_V $, $F_T $, $F_P $, and $F_A $ are the scalar, vector, tensor, pseudoscalar and pseudovector components of the amplitude, respectively; $s = (p_p^\mu   + p_j^\mu  ) = 4\varepsilon _{pc}^2 $ is the invariant energy, $q_c^2  =  - t =  - (p_p^\mu   - p_p^{\prime \mu}  )^2 $ is the invariant momentum transfer. The amplitude~(\ref{Eq7}) is a Lorentz invariant. In the $p$--$N$ c.m. frame it can be related to the amplitude~(\ref{Eq5}) in such a way that the following relationship is satisfied:~\cite{McN}
\begin{equation}
\bar u_p^{\mu '_1 } \left( {{\bf{k'}}_c } \right)\bar u_j^{\mu '_2 } \left( { - {\bf{k'}}_c } \right)\hat F_j \left( {s,q_c^2 } \right)u_p^{\mu _1 } \left( {{\bf{k}}_c } \right)u_j^{\mu _2 } \left( { - {\bf{k}}_c } \right) = \chi _p^{\mu '_1  + } \chi _j^{\mu '_2  + } f_j \left( {{\bf{q}}_c } \right)\chi _p^{\mu _1 } \chi _j^{\mu _2 } .
\label{Eq8}
\end{equation}

The covariant matrix element $\left\langle {{\bf{k'}}} \right|t^{(j)} \left| {\bf{k}} \right\rangle $ can be defined so that to have the following correspondence:
\begin{eqnarray}
\bar u_p^{\mu '_1 } \left( {{\bf{k'}}_c } \right)&\bar u_j^{\mu '_2 }&\left( { - {\bf{k'}}_c } \right)\left\langle {{\bf{k'}}_c } \right|t^{(j)} \left| {{\bf{k}}_c } \right\rangle u_p^{\mu _1 } \left( {{\bf{k}}_c } \right)u_j^{\mu _2 } \left( { - {\bf{k}}_c } \right) \nonumber \\
&&=4\varepsilon _{pc} \chi _p^{\mu '_1  + } \chi _j^{\mu '_2  + } \left\langle {{\bf{k'}}_c } \right|t^{(j)} \left| {{\bf{k}}_c } \right\rangle _{n{\mathop{\rm c}\nolimits} } \chi _p^{\mu _1 } \chi _j^{\mu _2 } \nonumber \\
&&=-16\pi \varepsilon _{pc} \bar u_p^{\mu '_1 } \left( {{\bf{k'}}_c } \right)\bar u_j^{\mu '_2 } \left( { - {\bf{k'}}_c } \right)\hat F_j \left( {s,q_c^2 } \right)u_p^{\mu _1 } \left( {{\bf{k}}_c } \right)u_j^{\mu _2 } \left( { - {\bf{k}}_c } \right) .
\label{Eq9}
\end{eqnarray}
In Eq.~(\ref{Eq9}) the matrix element in the left-hand side and the expression in the third line are invariant quantities. Thus, we may write the following covariant relation, which is valid in the $p$--$A$ c.m. frame:
\begin{equation}
\left\langle {{\bf{k'}}} \right|t^{(j)} \left| {\bf{k}} \right\rangle  =  - 16\pi \varepsilon _{pc} \gamma _p^0 \gamma _j^0 \hat F_j \left( {s,q^2 } \right) .
\label{Eq10}
\end{equation}
Let us emphasize that the left and right sides of Eq.~(\ref{Eq10}) are operators in the bispinor spaces of the proton and nucleon. Here, in the matrix element, we do not write explicitly the momenta of the nucleon. The matrix element of $t^{(j)} $ is obtained on the energy shell in the $p$--$A$ c.m. frame for the case of the elastic scattering, which corresponds to the situation $\left| {{\bf{k'}}} \right| = \left| {\bf{k}} \right| = k$ for the proton momenta. In the case of quasifree scattering, the nucleon in the nucleus must also have the same energy before and after the collision. Owing to these on-shell conditions of the energy and momentum conservation, the initial and final momenta of the projectile and struck nucleon should correspond to the situation of the Breit frame kinematics, with the nucleon initially moving relative to the c.m. of nucleus.\cite{McN,Wall3} Under these conditions, the magnitudes of the momentum transfers in the $p$--$N$ and $p$--$A$ c.m. frames coincide: $ - t = q_c^2  = q^2 $. Taking into account Eq.~(\ref{Eq10}) and supposing that the concrete choice of off-shell extrapolation of the matrix element is not very important when the matrix element decreases rapidly with the $q^2 $ increase,\cite{Wall1,Wall2,Ble2} we will take the following approximation for the matrix element of the operator $t_j \left( {{\bf{r}}_j } \right)$:
\begin{equation}
\left\langle {{\bf{k'}}} \right|t_j \left( {{\bf{r}}_j } \right)\left| {\bf{k}} \right\rangle  =  - 16\pi \varepsilon _{pc} \exp \left( {i{\bf{qr}}_j } \right)\gamma _p^0 \gamma _j^0 \hat F_j \left( {s,q^2 } \right) , \; {\bf{q}} = {\bf{k}} - {\bf{k'}} .
\label{Eq11}
\end{equation}
In the coordinate representation we have a local matrix element $\left\langle {{\bf{r'}}} \right|t_j \left( {{\bf{r}}_j } \right)\left| {\bf{r}} \right\rangle  = \delta ({\bf{r'}} - {\bf{r}})t_j \left( {{\bf{r}} - {\bf{r}}_j } \right)$ where
\begin{equation}
t_j \left( {\bf{r}} \right) =  - \frac{2}{{\pi ^2 }}\varepsilon _{pc} \gamma _p^0 \gamma _j^0 \int {d^3 q\exp ( - i{\bf{qr}})} \hat F_j \left( {s,q^2 } \right) .
\label{Eq12}
\end{equation}

When neglecting correlations between nucleons in the nucleus, one may present the ground state wave function of the nucleus $\Psi _0 $ in the form of a product of single-particle wave functions $\varphi _j ({\bf{r}}_j )$, which in the relativistic description of the nucleus\cite{Wale,Hor,Gam,Ser,Meng,Typ,Lal} are four-component bispinors $\varphi _j^{T,\alpha }  = (u_j^T ,w_j^T )$, where $u_j $ and $w_j $ are two-component spinors of the upper and lower (the latter being small in the standard representation) components. Let us introduce the single-nucleon densities of the nucleus, for a spinless nucleus the nonzero ones being only the scalar density $\rho _{Sj} $ and the time component of the vector density $\rho _{Vj} $:
\begin{equation}
\rho _{Sj} (r) = (\Psi _0^ +   \cdot \gamma _j^0 \delta ({\bf{r}} - {\bf{r}}_j )\Psi _0 ) = \frac{1}{{N_j }}\sum\limits_{i = 1}^{N_j } {\left[ {\left| {u({\bf{r}}_i )} \right|^2  - \left| {w({\bf{r}}_i )} \right|^2 } \right]} ,
\label{Eq13}
\end{equation}
\begin{equation}
\rho _{Vj} (r) = (\Psi _0^ +   \cdot \delta ({\bf{r}} - {\bf{r}}_j )\Psi _0 ) = \frac{1}{{N_j }}\sum\limits_{i = 1}^{N_j } {\left[ {\left| {u({\bf{r}}_i )} \right|^2  + \left| {w({\bf{r}}_i )} \right|^2 } \right]} , \; N_j  = N,Z .
\label{Eq14}
\end{equation}
Because there is no preferential direction, they depend on $r = |{\bf{r}}|$, while the space components of the vector density as well as all other densities vanish. Neglecting the nucleon correlations, we may insert the projection operator $\left| 0 \right\rangle \left\langle 0 \right|$ between neighbouring operators in Eq.~(\ref{Eq1}). As a result, in the propagators $\tilde G^{( + )} $ we have $H_t \left( {{\bf{\hat k}},\left\{ {{\bf{r}}_j } \right\}} \right) \to \sqrt {{\bf{\hat k}}^2  + M^2 } $ for the nucleus Hamiltonian, and every operator $t_j $ is averaged over $\Psi _0 $ in Eq.~(\ref{Eq4}):
\begin{eqnarray}
(\Psi _0^ +   \cdot t_j \left( {{\bf{r}} - {\bf{r}}_j } \right)\Psi _0 ) &=&  - \frac{2}{{\pi ^2 }}\varepsilon _{pc} \gamma _p^0 \int {d^3 q\;e^{ - i{\bf{qr}}} \int {d^3 r'\;} e^{i{\bf{qr'}}} } \nonumber \\
&&\times \left[ {\rho _{Sj} (r')F_{Sj} \left( {s,q^2 } \right) + \gamma _p^0 \rho _{Vj} (r')F_{Vj} \left( {s,q^2 } \right)} \right] .
\label{Eq15}
\end{eqnarray}
Further, considering the distinction between densities~(\ref{Eq13}) and~(\ref{Eq14}) caused by the contributions of the lower components as small (see, for example, Refs.~\citen{Hor,Gam}), at this stage we will assume these densities to be equal to each other in the formulae for the scattering amplitudes: $\rho _{Sj} (r) = \rho _{Vj} (r) \equiv \rho _0^{(j)} (r)$. As a result, we have:
\begin{equation}
(\Psi _0^ +   \cdot \left\langle {{\bf{r'}}} \right|t_j \left( {{\bf{r}}_j } \right)\left| {\bf{r}} \right\rangle \Psi _0 ) = \delta ({\bf{r'}} - {\bf{r}})\gamma _p^0 \tilde \tau _j \left( {\bf{r}} \right) ,
\label{Eq16}
\end{equation}
\begin{equation}
\tilde \tau _j \left( {\bf{r}} \right) =  - \frac{2}{{\pi ^2 }}\varepsilon _{pc} \int {d^3 q\;e^{ - i{\bf{qr}}} \left[ {F_{Sj} \left( {s,q^2 } \right) + \gamma _p^0 F_{Vj} \left( {s,q^2 } \right)} \right]} Q_0^{(j)} (q) ,
\label{Eq17}
\end{equation}
where the elastic nucleon formfactor has been introduced: $Q_0^{(j)} (q) = \int {d^3 r\;} e^{i{\bf{qr}}} \rho _0^{(j)} (r)$.

Further we shall consider the free propagator of the following form (we will simply denote $\gamma _p^0  \equiv \gamma ^0 $):
\begin{equation}
G^{( + )}  = \tilde G^{( + )} \gamma ^0  = \left[ {\gamma ^0 E - {\bm{\gamma }\bf{\hat k}} - m - \gamma ^0 \sqrt {{\bf{\hat k}}^2  + M^2 }  + i0} \right]^{ - 1} .
\label{Eq18}
\end{equation}
For the propagator~(\ref{Eq18}) we shall employ the eikonal expansion in powers of $1/k$, choosing the $z$-direction ${\bf{e}}_z \parallel {\bf{k}}_a $, where ${\bf{k}}_a  = ({\bf{k}}_i  + {\bf{k}}_f )/2$ is the average momentum for the $p$--$A$ scattering. Thus, we present the free propagator in the form: $G^{( + )}  \approx \varepsilon _t /(\varepsilon _p  + \varepsilon _t )\left[ {G_e^{( + )}  + \delta G_{ne}^{( + )} } \right]$, where the eikonal propagator and the noneikonal corrections of the first order have the following form in the  momentum representation:
\begin{equation}
G_e^{( + )}  =  - 2m\Lambda _ +  \left( {{\bf{k}}_\Lambda  } \right)\left[ {2{\bf{k}}_a \left( {{\bf{\hat k}} - {\bf{k}}_a } \right) - i0} \right]^{ - 1} ,
\label{Eq19}
\end{equation}
\begin{equation}
\delta G_{{\rm{ne}}}^{( + )}  = 2m\Lambda _ +  \left( {{\bf{k}}_\Lambda  } \right)\frac{{\left( {{\bf{\hat k}} - {\bf{k}}_a } \right)^2  - q^2 /4}}{{\left[ {2{\bf{k}}_a \left( {{\bf{\hat k}} - {\bf{k}}_a } \right) - i0} \right]^2 }} + \frac{{{\bm{\gamma }}({\bf{\hat k}} - {\bf{k}}_\Lambda  )}}{{2{\bf{k}}_a \left( {{\bf{\hat k}} - {\bf{k}}_a } \right) - i0}} .
\label{Eq20}
\end{equation}
Here, $\Lambda _ +  \left( {{\bf{k}}_\Lambda  } \right) = (\gamma ^0 \varepsilon _p  - {\bm{\gamma }\bf{k}}_\Lambda   + m)/(2m)$ is the projection operator onto the proton states with the positive energy and the momentum ${\bf{k}}_\Lambda  $. The direction of ${\bf{k}}_\Lambda  $ should be close to the chosen $z$-direction for the eikonal approximation, however we have a certain freedom in choosing it: we may take ${\bf{k}}_\Lambda   = {\bf{k}}_a $ or ${\bf{k}}_\Lambda   = {\bf{k}}_i $, or even ${\bf{k}}_\Lambda   = {\bf{k}}_f $ (we remind that ${\bf{k}}_i  = {\bf{k}}_a  + {\bf{q}}/2$, ${\bf{k}}_f  = {\bf{k}}_a  - {\bf{q}}/2$, ${\bf{q}} \bot {\bf{k}}_a $, ${\bf{q}} = {\bf{k}}_i  - {\bf{k}}_f $). Each of these variants is practically equivalent, although has its advantages when considering the eikonal approximation for the Dirac equation, and our choise of ${\bf{k}}_\Lambda  $ will be conctretized below from physical reasoning.

The eikonal propagator in the coordinate representation takes the form:
\begin{equation}
\left\langle {{\bf{r'}}} \right|G_e^{( + )} \left| {\bf{r}} \right\rangle  =  - \frac{i}{{2k_a }}2m\Lambda _ +  \left( {{\bf{k}}_\Lambda  } \right)\delta ({\bf{b'}} - {\bf{b}})\theta \left( {z' - z} \right)\exp \left[ {ik_a (z' - z)} \right] ,
\label{Eq21}
\end{equation}
where the transversal vector has been introduced: ${\bf{b}} = {\bf{r}}_ \bot  $, ${\bf{b}} \bot {\bf{e}}_z $. Neglecting the noneikonal corrections and restricting to the eikonal expressіon for the propagator~(\ref{Eq21}), with taking into account Eqs.~(\ref{Eq16}) and~(\ref{Eq17}), we find the amplitude of the elastic $p$--$A$ scattering~(\ref{Eq4}) in the form: $F\left( {{\bf{k}}_f ,{\bf{k}}_i } \right) = {\bar u^{\mu '} \left( {{\bf{k}}_f } \right) \tilde F_e ({\bf{q}})u^\mu  \left( {{\bf{k}}_i } \right)}$ where
\begin{eqnarray}
\tilde F_e \left( {\bf{q}} \right) &=& - \frac{1}{{4\pi }}\int {d^2 b\;e^{i{\bf{qb}}} \left\{ {\int {dz\sum\limits_{j = 1}^A {\hat \tau _j \left( {{\bf{b}},z} \right)} } } \right. } \nonumber \\
&&+ \sum\limits_{n = 2}^A {\left( {{\textstyle{\frac{- i}{2k_a }}}} \right)^{n - 1}} {\sum\limits_{i_1  = 1}^A \ldots\sum\limits_{i_n  \ne ...}^A {\int {dz_{i_1 }  \ldots } \,} dz_{i_n } \hat \tau _{i_1} \left( {{\bf{b}},z_{i_1 } } \right)} \nonumber \\
&&\times 2m\Lambda _+  \left( {{\bf{k}}_\Lambda  } \right) \hat \tau _{i_2 } \left( {{\bf{b}},z_{i_2 } } \right)2m\Lambda _ +  \left( {{\bf{k}}_\Lambda  } \right) \cdots \hat \tau _{i_{n - 1} } \left( {{\bf{b}},z_{i_{n - 1} } } \right) \nonumber \\
&&\left. \times 2m\Lambda _ +  \left( {{\bf{k}}_\Lambda  } \right) \hat \tau _{i_n } \left( {{\bf{b}},z_{i_n } } \right)\theta \left( {z_{i_1 }  - z_{i_2 } } \right) \ldots \theta \left( {z_{i_{n - 1} }  - z_{i_n } } \right) \vphantom{\sum\limits_{j = 1}^A} \right\} ,
\label{Eq22}
\end{eqnarray}
%\begin{equation}
\begin{eqnarray}
\hat \tau _j \left( {{\bf{b}},z} \right) &=&  - \frac{{\varepsilon _t \varepsilon _{pc} }}{{\varepsilon _p  + \varepsilon _t }}\frac{2}{{\pi ^2 }}\int {d^3 q}\;e^{ - i({\bf{qb}} + q_z z)}  \nonumber \\
&& \times \left[ F_{Sj} \left( {s,q^2 } \right) + \gamma ^0 F_{Vj} \left( {s,q^2 } \right) \right] Q_0^{(j)} (q) .
\label{Eq23}
\end{eqnarray}
%\end{equation}
In Eq.~(\ref{Eq22}) all indices $i_1  \ldots i_n $ in the sums have different values. Further we shall present the amplitude in terms of the matrix elements with respect to the spinors $\chi $ in the initial and final rest frames of the proton. Introducing the projection operator onto the upper components $B_ +   = (1 + \gamma ^0 )/2$ and the Lorentz boost operators for the proton:
\begin{equation}
L({\bf{k}}) = \sqrt {\frac{{\varepsilon _p  + m}}{{2m}}} \left[ {1 + \frac{{{\bm{\alpha }\bf{k}}}}{{\varepsilon _p  + m}}} \right] ,
\label{Eq24}
\end{equation}
we obtain the scattering amplitude in the form $F\left( {{\bf{k}}_f ,{\bf{k}}_i } \right) = {\chi ^{\mu ' + } \hat F_e ({\bf{q}})\chi ^\mu}$ where
\begin{eqnarray}
\hat F_e ({\bf{q}}) &=& 2mB_+  L( - {\bf{k}}_f )\tilde F_e \left( {\bf{q}} \right)L({\bf{k}}_i )B_+ = - \frac{1}{{4\pi }}\int {d^2 b\;} e^{i{\bf{qb}}} 2mB_+  L( -{\bf{k}}_f )  \nonumber \\
&& \times \left\{ {\int {dz\sum\limits_{j = 1}^A {\hat \tau _j \left( {{\bf{b}},z} \right)} } } \right. + \sum\limits_{n = 2}^A {\left( {{\textstyle{\frac{-i}{2k_a }}}} \right)^{n - 1} } \sum\limits_{i_1  = 1}^A  \ldots  \sum\limits_{i_n  \ne ...}^A {\int {dz_{i_1 }  \ldots } \,} dz_{i_n } \nonumber \\
&& \times \hat \tau _{i_1 } \left( {{\bf{b}},z_{i_1 } } \right)2m\Lambda _ +  \left( {{\bf{k}}_\Lambda  } \right)\hat \tau _{i_2 } \left( {{\bf{b}},z_{i_2 } } \right)2m\Lambda _ +  \left( {{\bf{k}}_\Lambda  } \right) \cdots \hat \tau _{i_{n - 1} } \left( {{\bf{b}},z_{i_{n - 1} } } \right) \nonumber \\
&& \left. {\times 2m\Lambda _+  \left( {{\bf{k}}_\Lambda  } \right)\hat \tau _{i_n } \left( {{\bf{b}},z_{i_n } } \right)\theta \left( {z_{i_1 }  - z_{i_2 } } \right) \ldots \theta \left( {z_{i_{n - 1}}- z_{i_n} } \right)} \vphantom{\sum\limits_{j = 1}^A}\right\}L({\bf{k}}_i )B_+ .
\label{Eq25}
\end{eqnarray}
For considering further calculations we write down the following useful formulae:
\begin{equation}
\Lambda _ +  \left( {{\bf{k}}_{i,f} } \right)\Lambda _ +  \left( {{\bf{k}}_{i,f} } \right) = \Lambda _ +  \left( {{\bf{k}}_{i,f} } \right) , \quad \Lambda _ +  \left( {{\bf{k}}_{i,f} } \right)\gamma ^0 \Lambda _ +  \left( {{\bf{k}}_{i,f} } \right) = \frac{{\varepsilon _p }}{m}\Lambda _ +  \left( {{\bf{k}}_{i,f} } \right) ,
\label{Eq26}
\end{equation}
\begin{equation}
\Lambda _+  \left( {{\bf{k}}_{i,f}} \right) = L\left( {{\bf{k}}_{i,f}} \right)B_+  L\left( { - {\bf{k}}_{i,f}} \right) ,
\label{Eq27}
\end{equation}
%\begin{equation}
\begin{eqnarray}
B_+ L\left( {-{\bf{k}}_f } \right)L\left( {{\bf{k}}_i } \right)B_+ &=& \frac{1}{{2m}}\left[ \varepsilon _p (1 - \cos \theta ) \right. \nonumber \\
&&\left. + m\;(1 + \cos \theta ) + i(\varepsilon _p  - m){\bf{\sigma n}}\;\sin \theta  \right] ,
\label{Eq28}
\end{eqnarray}
%\end{equation}
%\begin{equation}
\begin{eqnarray}
B_+ L\left( { - {\bf{k}}_f } \right)\gamma ^0 L\left( {{\bf{k}}_i } \right)B_+ &=& \frac{1}{{2m}}\left[ \varepsilon _p (1 + \cos \theta ) \right. \nonumber \\
&& \left. + m\;(1 - \cos \theta ) - i(\varepsilon _p  - m){\bf{\sigma n}}\;\sin \theta \right] ,
\label{Eq29}
\end{eqnarray}
%\end{equation}
\begin{equation}
B_+ L\left({-{\bf{k}}_{i,f}}\right)\gamma ^0 L\left({{\bf{k}}_{i,f}}\right)B_+ = \frac{{\varepsilon _p}}{m} , \quad L\left({-{\bf{k}}_{i,f}}\right)L\left({{\bf{k}}_{i,f}} \right) = 1 ,
\label{Eq30}
\end{equation}
where ${\bf{n}} = {\bf{k}}_i  \times {\bf{k}}_f /|{\bf{k}}_i  \times {\bf{k}}_f | $ is the normal to the scattering plane and $\theta $ is the scattering angle in the $p$--$A$ c.m. frame.

Using the representation~(\ref{Eq27}) for the projection operators $\Lambda _+ $ in Eq.~(\ref{Eq25}) and taking into account formulae~(\ref{Eq28})--(\ref{Eq30}), we see that, when choosing ${\bf{k}}_\Lambda = {\bf{k}}_a $ in all $\Lambda _+ $, we obtain a $T$-invariant expression for the amplitude but in the multiple scattering terms there are only two $pN$-interaction operators with spin rotation (i.e. those containing the operator ${\bm{\sigma}\bf{n}}$), namely, at the first and last collisions. Taking ${\bf{k}}_\Lambda = {\bf{k}}_i $ or ${\bf{k}}_\Lambda = {\bf{k}}_f $ in all $\Lambda _+ $, we have only one operator with spin rotation at the last or at the first collision, correspondingly, and besides, the model will not be $T$-invariant. A more reasonable model would be the one in which the operator with the spin rotation could appear in each of the $n$ successive collisions. In order to obtain such a picture as well as the $T$-invariance of the model, in the expression for the amplitude~(\ref{Eq25}) we shall simultaneously use the projection operators with ${\bf{k}}_\Lambda = {\bf{k}}_i $ and ${\bf{k}}_\Lambda = {\bf{k}}_f $, writing Eq.~(\ref{Eq25}) in the following symmetrized form:
\begin{eqnarray}
\hat F_e ({\bf{q}}) &=& -\frac{1}{{4\pi }}\int {d^2 b\;} e^{i{\bf{qb}}} 2mB_+  L( -{\bf{k}}_f )\left\{ {\int {dz\sum\limits_{j = 1}^A {\hat \tau _j \left( {{\bf{b}},z} \right)}}} \right. + \sum\limits_{n = 2}^A {\left( {{\textstyle{\frac{-i}{2k_a}}}} \right)^{n-1}} \nonumber \\
&& \times \frac{1}{n}\sum\limits_{l=1}^n {\sum\limits_{i_1  = 1}^A  \ldots  } \sum\limits_{i_n  \ne ...}^A {\int {dz_{i_1}  \ldots } \,} dz_{i_n } \hat \tau _{i_1 } \left( {{\bf{b}},z_{i_1 } } \right) 2m\Lambda _+  \left( {{\bf{k}}_1 (l)} \right) \nonumber \\
&& \times \hat \tau _{i_2 } \left( {{\bf{b}},z_{i_2 } } \right)2m\Lambda _ +  \left( {{\bf{k}}_2 (l)} \right) \cdots \hat \tau _{i_{n - 1} } \left( {{\bf{b}},z_{i_{n - 1} } } \right) 2m\Lambda _+ \left( {{\bf{k}}_{n - 1} (l)} \right) \nonumber \\
&& \times   \left. { \hat \tau _{i_n } \left( {{\bf{b}},z_{i_n } } \right)\theta \left( {z_{i_1 } -z_{i_2}} \right) \ldots \theta \left( {z_{i_{n-1}} - z_{i_n } } \right)} \vphantom{\sum\limits_{j = 1}^A} \right\}L({\bf{k}}_i )B_+ ,
\label{Eq31}
\end{eqnarray}
%(31)
where the momenta in the operators $\Lambda _+ $ are as follows: ${\bf{k}}_m (l) = {\bf{k}}_f
$ for $1 \le m < l$, and ${\bf{k}}_m (l) = {\bf{k}}_i $ for $l \le m < n - 1$, $1 \le l \le n$. Using the representation~(\ref{Eq27}) for the operators $\Lambda _+ $ and formulae~(\ref{Eq28})--(\ref{Eq30}), we obtain the following expression:
\begin{eqnarray}
\hat F_e ({\bf{q}}) &=&  - \frac{1}{{4\pi }}\int {d^2 b\;} e^{i{\bf{qb}}} \sum\limits_{n = 1}^A {\left( {{\textstyle{\frac{-i}{2k_a }}}} \right)^{n - 1} } \frac{1}{n}\sum\limits_{\{ i_1 ...i_n \} } {\sum\limits_{i_l  = i_1 }^{i_n } {\int\limits_{ - \infty }^\infty  {dz_1  \ldots dz_n } } } \nonumber \\
&& \times \hat \tau '_{i_l } \left( {{\bf{b}},z_l } \right)\bar \tau _{i_1 } \left( {{\bf{b}},z_1 } \right)... \bar \tau _{i_{l - 1} } \left( {{\bf{b}},z_{l - 1} } \right)\bar \tau _{i_{l + 1} } \left( {{\bf{b}},z_{l + 1} } \right) \cdots \bar \tau _{i_n } \left( {{\bf{b}},z_n } \right) ,
\label{Eq32}
\end{eqnarray}
%\begin{equation}
\begin{eqnarray}
\hat \tau '_{i_l } \left( {{\bf{b}},z} \right) &=& B_+ L\left({-{\bf{k}}_f} \right)\hat \tau _{i_l } \left( {{\bf{b}},z} \right)L\left( {{\bf{k}}_i } \right)B_+ ,  \nonumber \\
\bar \tau _{i_m } \left( {{\bf{b}},z} \right) &=& B_+ L\left({ -{\bf{k}}_{i,f}} \right)\hat \tau _{i_m } \left( {{\bf{b}},z} \right)L\left( {{\bf{k}}_{i,f} } \right)B_+ .
\label{Eq33}
\end{eqnarray}
%\end{equation}
Here, $\sum\nolimits_{\{ i_1 ...i_n \} } {}$ is the sum over possible combinations of indices $i_1 ...i_n $ . Obtaining Eq.~(\ref{Eq32}), we have taken into account the commutativity of all functions $\hat \tau '_{i_l } \left( {{\bf{b}},z} \right)$ and $\bar \tau _{i_m } \left( {{\bf{b}},z} \right)$, which allowed us to sum up the $\theta $-functions in Eq.~(\ref{Eq31}). Integrating in Eq.~(\ref{Eq32}) over all $z$-variables and collecting first the terms with the same values $i_l $ and $i_1 ...i_n  \ne i_l $ and then the terms of the same sort, we find the following expression for the amplitude of the elastic $p$--$A$ scattering:
\begin{eqnarray}
\hat F_e ({\bf{q}}) &=& \frac{{ik}}{{2\pi }}\int {d^2 b\;} e^{i{\bf{qb}}} \left\{ {Z\hat E'_p (b)} \sum\limits_{n_p  = 0}^{Z - 1}\right. {\sum\limits_{n_n  = 0}^N {\frac{(-1)^{n_p +n_n }}{{n_p  + n_n  + 1}}C_{n_p }^{Z - 1} C_{n_n }^N } } \bar E_p^{n_p } (b)\bar E_n^{n_n } (b) \nonumber \\
&& + N\hat E'_n (b) \left. \sum\limits_{n_p  = 0}^Z {\sum\limits_{n_n  = 0}^{N - 1} {\frac{( - 1)^{n_p +n_n }}{{n_p  + n_n  + 1}}C_{n_p }^Z C_{n_n }^{N - 1} } } \bar E_p^{n_p } (b){\bar E_n^{n_n } (b)} \right\} ,
\label{Eq34}
\end{eqnarray}
where $C_{n_p }^Z $ are the binomial coefficients and the following functions for nucleons have been introduced:
\begin{eqnarray}
\bar E_j \left( b \right) &\equiv& \frac{i}{{2k_a }}\int\limits_{ - \infty }^\infty  {dz} \;\bar \tau _j \left( {{\bf{b}},z} \right)
= \frac{2}{{ik_a \pi }}\frac{{2\varepsilon _t \varepsilon _{pc} }}{{(\varepsilon _p  + \varepsilon _t )m}}\int {d^2 q}\;e^{ - i{\bf{qb}}} \nonumber \\
&& \times 2m\left[ {mF_{Sj} \left({s,q^2}\right) + \varepsilon _p F_{Vj} \left({s,q^2}\right)} \right] Q_0^{(j)} (q) ,
\label{Eq35}
\end{eqnarray}
\begin{eqnarray}
\hat E'_j \left( b \right) &\equiv& \frac{i}{{2k_a }}\int\limits_{ - \infty }^\infty  {dz} \;\hat \tau '_j \left( {{\bf{b}},z} \right)
= \frac{1}{{ik_a \pi }}\frac{{2\varepsilon _t \varepsilon _{pc} }}{{(\varepsilon _p  + \varepsilon _t )m}}\int {d^2 q}\;e^{ - i{\bf{qb}}}
\nonumber \\
&& \times 2m \left\{ {(1 - \cos \theta )\left[ {\varepsilon _p F_{Sj}\left({s,q^2}\right)  + mF_{Vj}\left({s,q^2}\right)} \right]}\right. \nonumber \\
&& + (1 + \cos \theta )\left[ {mF_{Sj}\left({s,q^2}\right)  + \varepsilon _p F_{Vj}\left({s,q^2}\right)} \right] \nonumber \\
&&  + i(\varepsilon _p  - m){\bm{\sigma }\bf{ n}}\;\sin \theta \;\left. {\left[ {F_{Sj}\left({s,q^2}\right)  - F_{Vj}\left({s,q^2}\right)} \right]} \right\}Q_0^{(j)} (q) .
\label{Eq36}
\end{eqnarray}
The expression~(\ref{Eq34}) for the elastic $p$--$A$ scattering amplitude can also be represented in the form:
\begin{eqnarray}
\hat F_e ({\bf{q}}) &=& \frac{{ik}}{{2\pi }}\int {d^2 b\;} e^{i{\bf{qb}}} \int\limits_0^1 {dx} \left\{ {Z\hat E'_p } (b)\left[ {1 - x\bar E_p } \right]^{Z - 1} \left[ {1 - x\bar E_n } \right]^N  \right. \nonumber \\
&& + N\hat E'_n (b)\left[ {1 - x\bar E_p } \right]^Z \left. {\left[ {1 - x\bar E_n } \right]^{N - 1} } \right\} .
\label{Eq37}
\end{eqnarray}

Let us return to the definition of the $pN$-amplitudes~(\ref{Eq5})--(\ref{Eq8}). In calculations by MDST, it is usual to take into account  only the central and spin-orbit parts of the $pN$-amplitudes (see, for example, Refs.~\citen{Osl, Ble2}). Further we will also restrict our consideration to these amplitudes $A\left( q \right)$ and $C\left( q \right)$ in Eq.~(\ref{Eq5}). For this purpose, instead of the Eq.~(\ref{Eq8}) we employ a simplified relation between the amplitudes, similarly to the work of Ref.~\citen{Ble3}, which may be written in the $p$--$A$ c.m. frame with making use of the forward scattering conditions, as follows:
%\begin{equation}
\begin{eqnarray}
B_+ L( -{\bf{k}}_f )\left[ F_{Sj} \left({s,q^2}\right) \right. + &\gamma _p^0& \left. F_{Vj} \left({s,q^2}\right)\right] L({\bf{k}}_i ) B_+  \nonumber \\
&& = \frac{{k_c }}{{2mk}}\left[ {A_j (q) + qC_j (q){\bm{\sigma }\bf{ n}}} \right] .
\label{Eq38}
\end{eqnarray}
%\end{equation}
Strictly speaking, the used $pN$-amplitudes should refer to the scattering in the Breit frame, but here we will neglect this distinction from the usual transition between the $p$--$N$ c.m. and $p$--$A$ c.m. frames because it can yield considerable differences in the calculated $p$--$A$ scattering observables only for light target nuclei and for larger scattering angles.\cite{McN} From Eq.~(\ref{Eq38}) we obtain the explicit formulae for the relationship between the $pN$-amplitudes:
\begin{equation}
2mF_{Sj} \left( {s,q^2 } \right) = \frac{k}{{k_c }}\left\{ {\frac{A_j (q)}{{2(\varepsilon _p  + m)}} - \frac{iC_j (q)}{{\sqrt {k^2  - q^2 /4} }}\left[ {\varepsilon _p  - \frac{{q^2 }}{{4(\varepsilon _p  + m)}}} \right]} \right\} ,
\label{Eq39}
\end{equation}
\begin{equation}
2mF_{Vj} \left( {s,q^2 } \right) = \frac{k}{{k_c }}\left\{ {\frac{A_j (q)}{{2(\varepsilon _p  + m)}} + \frac{iC_j (q)}{{\sqrt {k^2  - q^2 /4} }}\left[ {m + \frac{{q^2 }}{{4(\varepsilon _p  + m)}}} \right]} \right\} .
\label{Eq40}
\end{equation}
We write the amplitudes in the $p$--$N$ c.m. frame in the form: $A\left( q \right) = ik_c /(2\pi )f_c (q)$, $qC\left( q \right) = ik_c /(2\pi )f_s (q)$. Taking into account that $2\varepsilon _t \varepsilon _{pc} /((\varepsilon _p  + \varepsilon _t )m) \approx k/k_c $, the functions~(\ref{Eq35}) and~(\ref{Eq36}) can be written as follows:
\begin{equation}
\bar E_j \left( b \right) = E_0^{(j)} \left( b \right) + \bar E_s^{(j)} \left( b \right) ,  \;
\hat E'_j \left( b \right) = \bar E_j \left( b \right) + \left[(1-\cos \theta ) + i{\bm{\sigma }\bf{ n}}\;\sin \theta \right] E_s^{\prime (j)} \left( b \right) ,
\label{Eq41}
\end{equation}
\begin{equation}
E_0^{(j)} \left( b \right) = \frac{1}{{2\pi }}\int\limits_0^\infty  {dqq\;J_0 (qb)f_c^{(j)} (q)} Q_0^{(j)} (q) ,
\label{Eq42}
\end{equation}
\begin{equation}
\bar E_s^{(j)} \left( b \right) = \frac{i}{{4\pi }}\int\limits_0^\infty  {dq\;J_0 (qb)\frac{{q^2 }}{{\sqrt {k^2  - q^2 /4} }}f_s^{(j)} (q)} Q_0^{(j)} (q) ,
\label{Eq43}
\end{equation}
\begin{equation}
E_s^{\prime (j)} \left( b \right) =  - \frac{i}{{2\pi }}\int\limits_0^\infty  {dq\;J_0 (qb)\frac{{k^2 }}{{\sqrt {k^2  - q^2 /4} }}f_s^{(j)} (q)} Q_0^{(j)} (q) .
\label{Eq44}
\end{equation}

Thus, the $pA$-amplitude in Eq.~(\ref{Eq37}) can be represented in the form: $\hat F_e ({\bf{q}}) = A_0 (q) + {\bm{\sigma}\bf{n}}B_0 (q)$ where
\begin{equation}
A_0 (q) = ik\int\limits_0^\infty  {dbb\;J_0 (qb)\Omega _A (b)} ,  \quad
B_0 (q) =  - k\sin \theta \int\limits_0^\infty  {dbb\;J_0 (qb)\tilde \Omega _B (b)} ,
\label{Eq45}
\end{equation}
and the nuclear profile functions introduced in Eq.~(\ref{Eq45}) are equal to
\begin{eqnarray}
\Omega _A (b) &=& \int\limits_0^1 {dx} \left\{ {Z\left[ {\bar E_p (b) + (1-\cos \theta )E_s^{\prime (p)} \left( b \right)} \right]} \right.\left[ {1 - x\bar E_p (b)} \right]^{Z - 1} \left[ {1 - x\bar E_n (b)} \right]^N  \nonumber \\
&& + N\left[ {\bar E_n (b) + (1-\cos \theta )E_s^{\prime (n)} \left( b \right)} \right]\left[ {1 - x\bar E_p (b)} \right]^Z \left. {\left[ {1 - x\bar E_n (b)} \right]^{N - 1} } \right\} ,
\label{Eq46}
\end{eqnarray}
%\begin{equation}
\begin{eqnarray}
\tilde \Omega _B (b) &=& \int\limits_0^1 {dx} \left\{ {ZE_s^{\prime (p)} \left( b \right)\left[ {1 - x\bar E_p } \right]^{Z - 1} } \right.\left[ {1 - x\bar E_n } \right]^N  \nonumber \\
&& + NE_s^{\prime (n)} \left( b \right)\left[ {1 - x\bar E_p } \right]^Z \left. {\left[ {1 - x\bar E_n } \right]^{N - 1} } \right\} .
\label{Eq47}
\end{eqnarray}
%\end{equation}
If the proton and neutron $E$-functions may be put identical (for example, taken in the averaged form), then the expressions~(\ref{Eq46})  and~(\ref{Eq47}) are simplified:
%\begin{equation}
\begin{eqnarray}
\Omega _A (b) &=& \left[ {1 + (1 - \cos \theta )\frac{{E'_s \left( b \right)}}{{\bar E(b)}}} \right]\left\{ {1 - \left[ {1 - \bar E(b)} \right]^A } \right\} ,  \nonumber \\
\tilde \Omega _B (b) &=& \frac{{E'_s \left( b \right)}}{{\bar E(b)}}\left\{ {1 - \left[ {1 - \bar E(b)} \right]^A } \right\} .
\label{Eq48}
\end{eqnarray}
%\end{equation}

When taking account of the electromagnetic effects, the central $A_0 \left( q \right)$ and spin-orbit $B_0 \left( q \right)$ amplitudes of the elastic $p$--$A$ scattering can be written as:
%\begin{equation}
\begin{eqnarray}
A_0 (q) &=& A_C \left( q \right) + ik\int\limits_0^\infty  {db} bJ_0 (qb)\left\{ \left( {e^{i\chi _0 \left( b \right)}  - e^{i\chi _1 \left( b \right)} } \right) \right. \nonumber \\
&& \left.+ e^{i\chi _1 \left( b \right)} \left[ {\Omega _A \left( b \right) - \Omega _B \left( b \right)\chi _{1s} (b)} \right] \right\} ,
\label{Eq49}
\end{eqnarray}
%\end{equation}
\begin{eqnarray}
B_0 (q) &=& B_C \left( q \right) - ik\int\limits_0^\infty  {db} bJ_1 (qb)\left\{ {\left[ {e^{i\chi _0 \left( b \right)} \chi _{0s} \left( b \right) - e^{i\chi _1 \left( b \right)} \chi _{1s} \left( b \right)} \right]} \right. \nonumber \\
&& \left. + e^{i\chi _1 \left( b \right)}{\left[ {\Omega _A \left( b \right)\chi _{1s} \left( b \right) + \Omega _B \left( b \right)} \right]} \right\} .
\label{Eq50}
\end{eqnarray}
where we have introduced the nuclear spin-orbit profile function $\Omega _B \left( b \right) = (i/k)d\tilde \Omega _B (b)/db $. In Eqs.~(\ref{Eq49}) and~(\ref{Eq50}), $\chi _0 \left( b \right) = 2\xi ln\left( {kb} \right)$ and $\chi _{0s} \left( b \right) = {{2\xi \kappa } \mathord{\left/ {\vphantom {{2\xi \kappa } b}} \right. \kern-\nulldelimiterspace} b}$ are the central and spin-orbit scattering phases in the Coulomb field of two point charges in the eikonal approximation. The central and spin-orbit components of the scattering amplitude, corresponding to these phases, are given by the formulae
\begin{equation}
A_C \left( q \right) =  - \frac{{2\xi k}}{{q^2 }}\frac{{\Gamma \left( {1 + i\xi } \right)}}{{\Gamma \left( {1 - i\xi } \right)}}\exp\left( { - 2i\xi \ln\frac{q}{{2k}}} \right) , \quad
B_C \left( q \right) =  - i\kappa qA_C \left( q \right) .
\label{Eq51}
\end{equation}
%(51)
Here, $\xi  = {{Ze^2 } \mathord{\left/ {\vphantom {{Ze^2 } {\hbar v}}} \right. \kern-\nulldelimiterspace} {\hbar v}}$ is the Sommerfeld parameter for the $p$--$A$ scattering, and the parameter $\kappa$ characterizes the spin-orbit quantities $\chi _{0s} \left( b \right)$ and $B_C \left( q \right)$. The eikonal Coulomb phase $\chi _1 \left( b \right)$ of the scattering on the volume charge of the target nucleus has the following form:~\cite{Ahm}
\begin{equation}
\chi _1 \left( b \right) = \chi _0 \left( b \right) + 8\pi \xi \int\limits_b^\infty  {drr^2 \rho _0^{\left( p \right)} \left( r \right)\left[ {\ln\left( {\frac{{1 + \sqrt {1 - {{b^2 } \mathord{\left/ {\vphantom {{b^2 } {r^2 }}} \right. \kern-\nulldelimiterspace} {r^2 }}} }}{{{b \mathord{\left/ {\vphantom {b r}} \right. \kern-\nulldelimiterspace} r}}}} \right) - \sqrt {1 - {{b^2 } \mathord{\left/  {\vphantom {{b^2 } {r^2 }}} \right.  \kern-\nulldelimiterspace} {r^2 }}} } \right]} ,
\label{Eq52}
\end{equation}
%(52)
and the corresponding spin-orbit phase is $\chi _{1s} \left( b \right) = (\kappa k)d\chi _1 \left( b \right)/db$, which is analogous to the usual macroscopic allowance for the interaction of the proton magnetic moment with the electromagnetic field of the target nucleus in MDST\cite{Osl} by introducing a spin-orbit correction to the macroscopic proton--nucleus Coulomb phase shift $\chi _1 \left( b \right)$. However, in contrast to this, in our previous works in the MDST framework we considered this interaction microscopically through including the Coulomb spin-orbit term in the proton--nucleon amplitudes. In this case, the value of parameter $\kappa$ was determined from the asymptotic behavior of the profile function $\Omega _B \left( b \right)$ at $b \to \infty $ (see, for example, Refs.~\citen{Kup2,Ber}), which leads to the value $\kappa  = \kappa _{pp} k_c /k$ where $\kappa _{pp} $ is the corresponding spin-orbit parameter in the $pp$-amplitude. According to concrete conventions of a used phase analysis, it may be $\kappa _{pp}  = (3 + 4\mu _a )/4m^2 $ or $\kappa _{pp}  = (3/(E_c  + m) + 2\mu _a /m)/2E_c $, where $\mu _a  = 1.79$ is the anomalous magnetic moment of the proton. This $\kappa$ value differs from that used in Ref.~\citen{Osl}. In order to ensure the correspondence with our microscopic consideration in Ref.~\citen{Kup2}, here we shall assume $\kappa  = \kappa _{pp} k_c /k$ with the latter variant of $\kappa _{pp} $ in accordance with the used phase analysis for the $pN$-amplitude.

\section{Results of calculations of the $p$--$A$ scattering observables}

To perform the calculations of the $p$--$A$ scattering amplitudes in the framework of the multiple scattering model, it is necessary to know the densities of nucleon distribution in the target nucleus. In this work, we have employed the nucleon densities obtained by us from the microscopic calculations of the nuclear structure in the approximation of the relativistic mean field (RMF). The RMF models known from the literature provide the description of properties of the ground state of finite nuclei with a good accuracy (see Refs.~\citen{Gam,Lal}).

The RMF model used in the present work is based on the nucleus Lagrangian density which has the following form:\cite{Typ,Lal}
%\begin{array}{l}
\begin{eqnarray}
{\cal L} &=& \sum\limits_{j = n,p} {\bar \Psi _j \left\{ {i\gamma ^\mu  \partial _\mu   - m_N  + g_\sigma  \sigma  - g_\omega  \gamma ^\mu  \omega _\mu   - \frac{{g_\rho  }}{2}\gamma ^\mu  {\bm{\rho }}_\mu   \cdot {\bm{\tau }} - \frac{e}{2}\gamma ^\mu  (1 - \tau _3 )A_\mu  } \right\}\Psi _j }  +  \nonumber \\
&& + \frac{1}{2}\partial ^\mu  \sigma \partial _\mu  \sigma  - \frac{1}{2}m_\sigma ^2 \sigma ^2  - \frac{1}{4}\Omega ^{\mu \nu } \Omega _{\mu \nu }  + \frac{1}{2}m_\omega ^2 \omega _\mu  \omega ^\mu  \nonumber \\
&& - \frac{1}{4}{\bf{R}}^{\mu \nu } {\bf{R}}_{\mu \nu }  + \frac{1}{2}m_\rho ^2 {\bm{\rho }}_\mu   \cdot {\bm{\rho }}^\mu  \; - \frac{1}{4}F^{\mu \nu } F_{\mu \nu } ,
\label{Eq53}
\end{eqnarray}
%\end{array}
%(53)
where $\sigma $, $\omega ^\mu  $ and ${\bm{\rho }}^\mu  $ are the scalar, isoscalar-vector and isovector-vector meson fields, respectively ($\mu  = 0,1,2,3$); $A_\mu  $ is the photon field ($e$ being the constant of electromagnetic interaction); $\Psi _{n,p} $ are the nucleon fields; ${\bm{\tau }}$ are the isospin Pauli matrices ($\tau _3  = 1$ for neutron, and $\tau _3  =  - 1$ for proton); $\gamma ^\mu  $ are the Dirac matrices; $g_\sigma  $, $g_\omega  $, and $g_\rho  $ are the meson--nucleon couplings, which depend on the nucleon density in the nucleus; $m_\sigma  $, $m_\omega $, $m_\rho  $, and $m_N $ are the masses of the mesons and nucleon; $\Omega _{\mu \nu } $, ${\bf{R}}_{\mu \nu } $, and $F_{\mu \nu } $ are the tensors of the fields of vector mesons and the electromagnetic field. All the meson--nucleon couplings and some of the meson masses are adjustable parameters of the model, whose values are determined in the literature from the requirement of the best description of properties of finite nuclei and nuclear matter. Owing to the stationarity of the problem and a number of symmetry requirements imposed when deriving the RMF equations, they involve only the time components of the four-vectors of the nucleon and electromagnetic currents and vector-meson fields. In the RMF models under consideration the antiparticle states are not taken into account. The corresponding set of coupled equations, which is presented in Refs.~\citen{Typ,Lal} and includes the Dirac equations for the spinor nucleon fields, the nonhomogeneous Klein--Gordon equations for the meson fields, and the equation for the Coulomb field, was solved by us numerically by the iteration method. In the calculations for the $^{40}$Ca nucleus, after obtaining the self-consistent solution, we took into account the center-of-mass motion by recalculating the neutron- and proton-density distributions in the harmonic-oscillator approximation as it was done in Ref.~\citen{Kup2}.

\begin{figure}[th]
\centerline{\includegraphics[width=5.2cm]{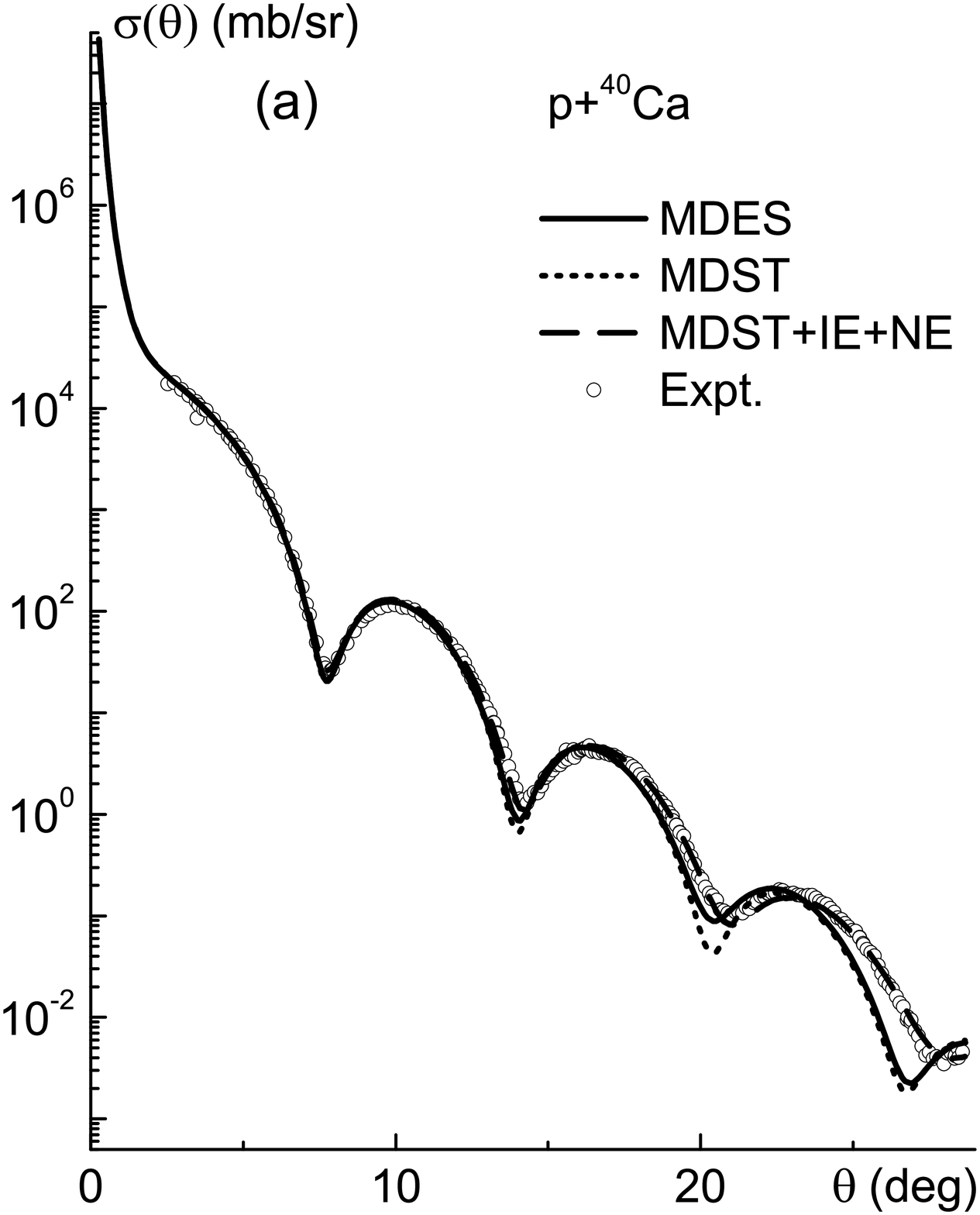}
%\;
\includegraphics[width=5.2cm]{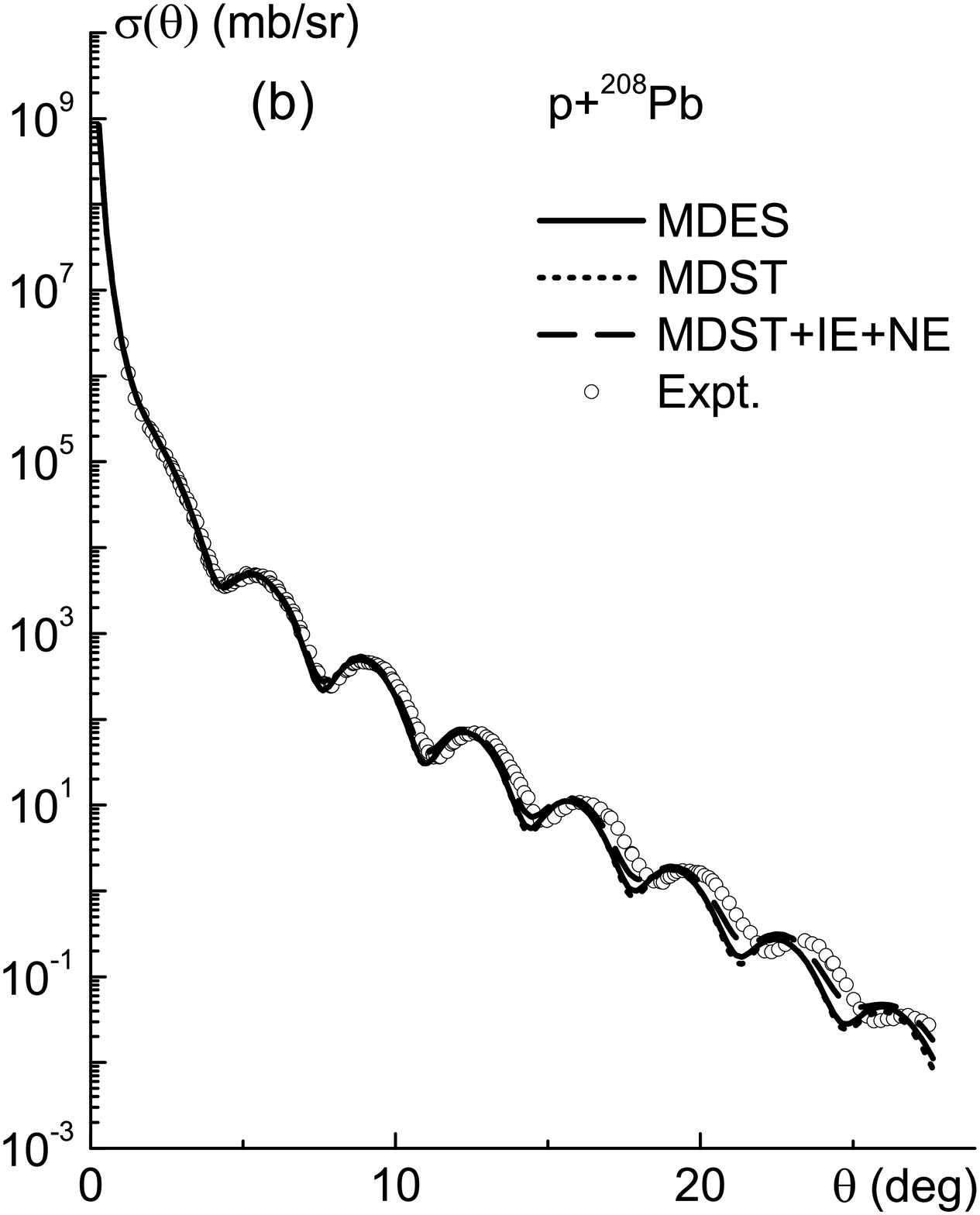}}

\medskip
\centerline{\includegraphics[width=5.2cm]{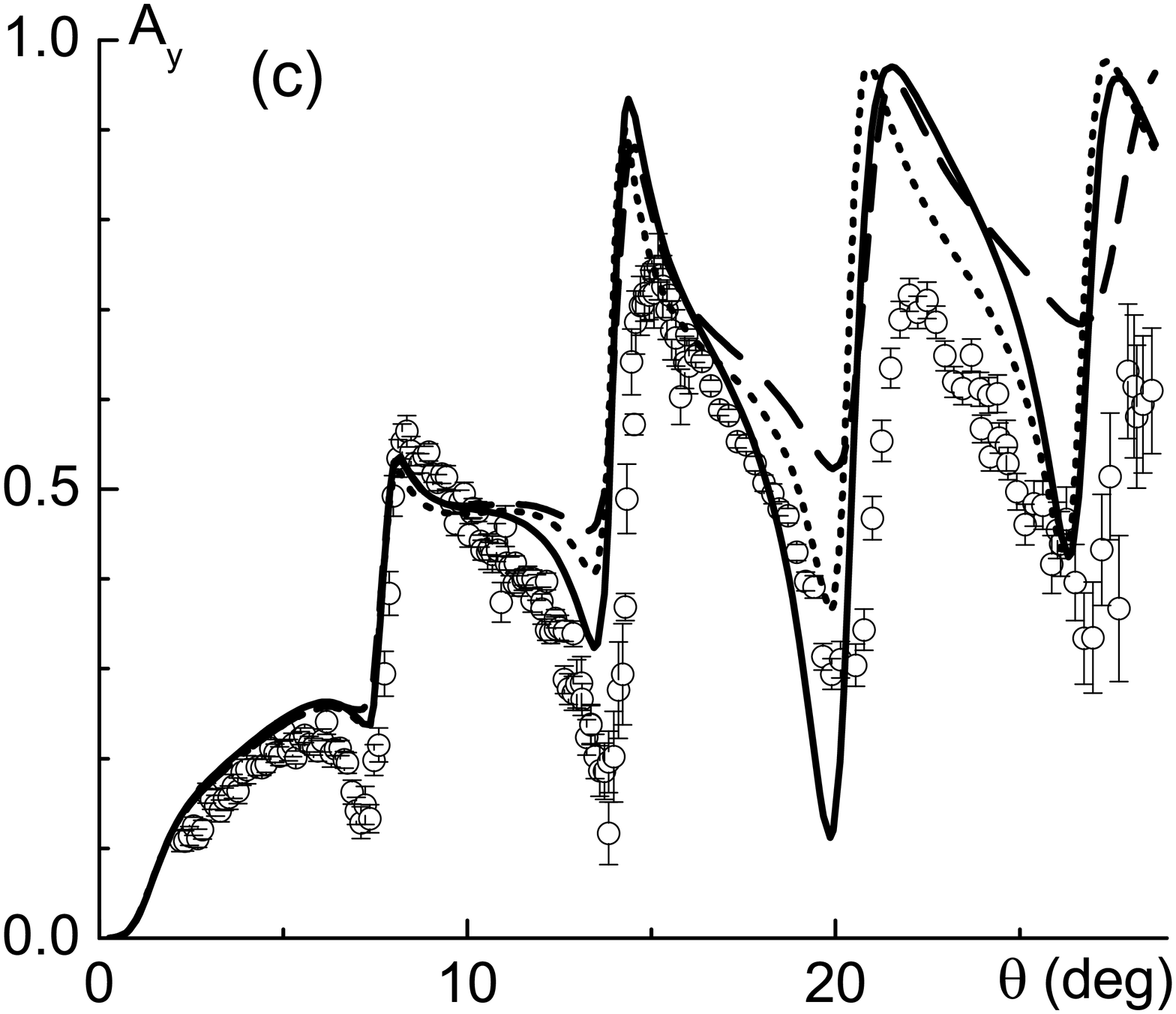}
%\;
\includegraphics[width=5.2cm]{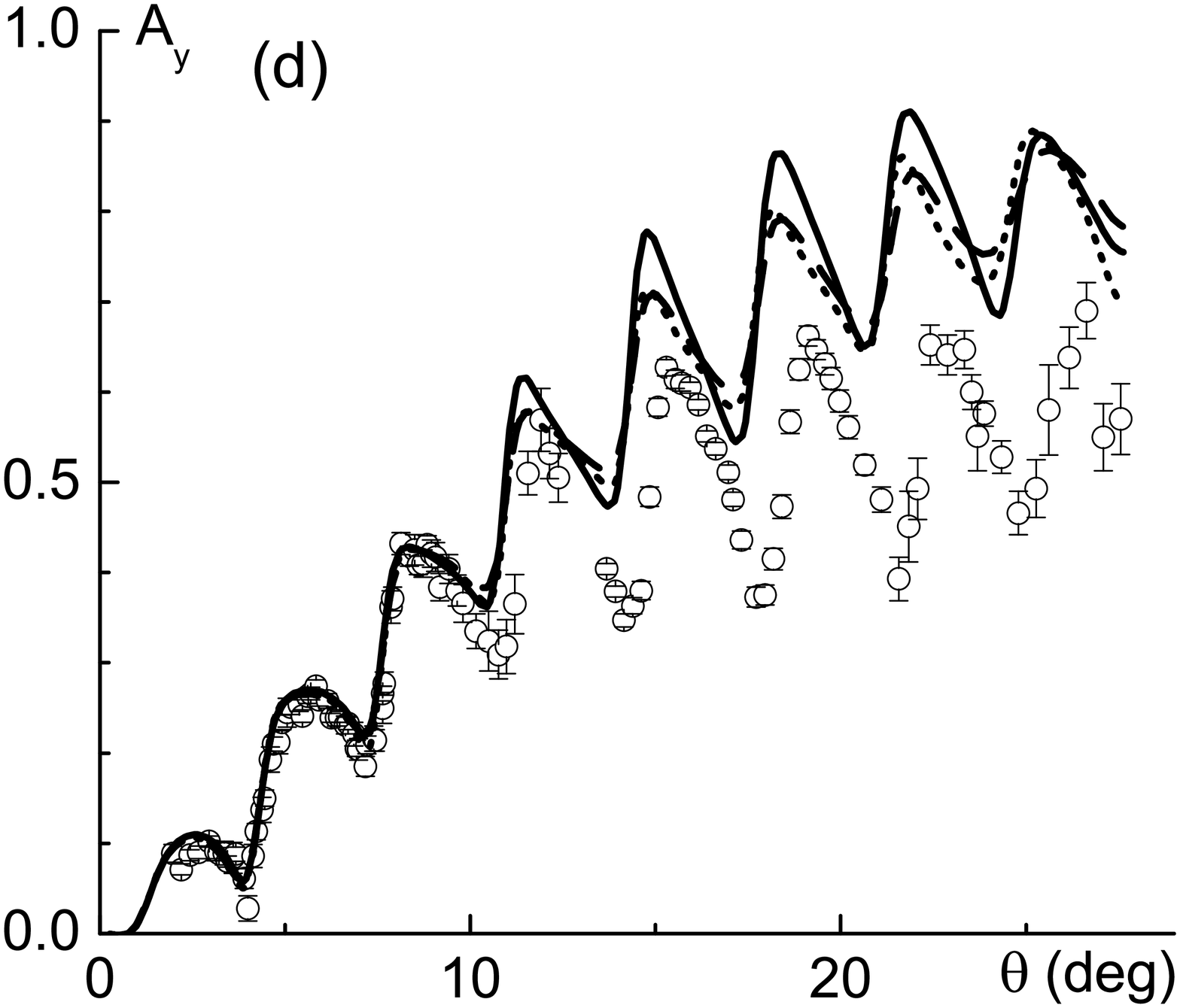}}

\medskip
\centerline{\includegraphics[width=5.2cm]{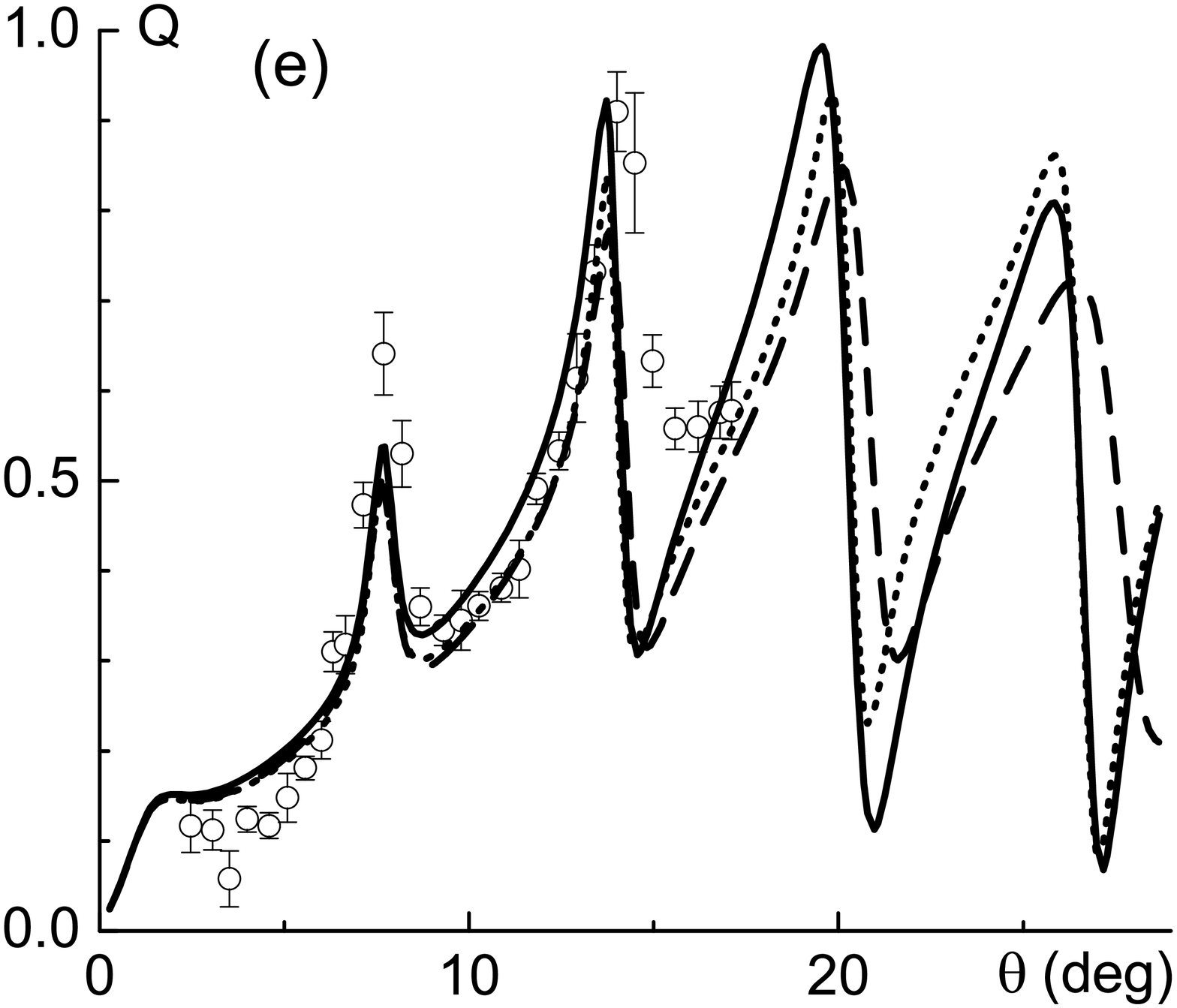}
%\;
\includegraphics[width=5.2cm]{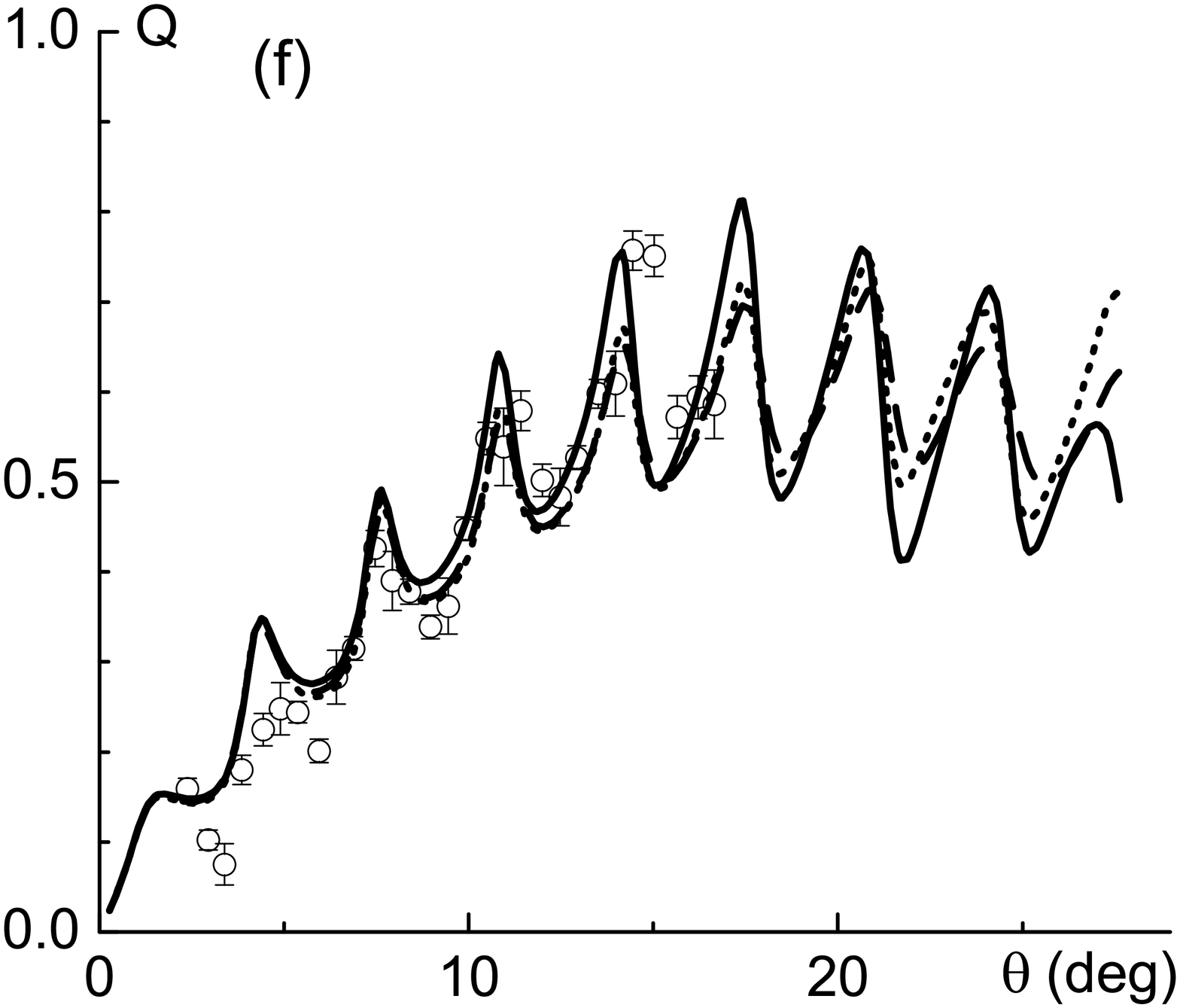}}
\caption{\label{F1}Differential cross sections $\sigma \left( \theta  \right)$, analyzing powers $A_y \left( \theta  \right)$, and spin rotation functions $Q\left( \theta  \right)$ for the elastic $p+^{40}$Ca and $p+^{208}$Pb scattering at 800 MeV, calculated with the RMF nucleon densities for the DD-ME2 interaction on the basis of the new MDES model in comparison with calculations by MDST without corrections as well as with the IE and NE corrections. The experimental data are taken from Refs.~\citen{Ble4,Hof1,Hof2,Ferg}.}
\end{figure}

By means of the MDES model developed above, which bases on the eikonal approximation for the Dirac equation and on employing realistic nucleon densities calculated in the RMF approximation and $pN$-amplitudes found from the phase-analysis solutions,  we have developed an original numerical computer code and performed the corresponding analysis of the differential cross sections, analyzing powers, and spin rotation functions for the elastic $p+^{40}$Ca and $p+^{208}$Pb scattering at the proton energy of 800 MeV with using different variants of the relativistic effective $NN$-interaction in the nuclear-structure calculations. The results obtained by us on the basis of MDES are compared with the analogous results obtained in the framework of the MDST approach under the same calculation conditions. In Fig.~\ref{F1}, we present the results of such calculations with making use of the nucleon densities, obtained by us in the RMF approach with the DD-ME2 interaction\cite{Lal} (note that in Ref.~\citen{ Kup2} a comparison was performed of the results of the MDST calculations with using different nucleon densities in the target nuclei obtained in the RMF and Skyrme--Hartree--Fock approaches). In these calculations we have used the approximation from Ref.~\citen{Kud} for the $pN$-amplitudes, which was determined from the phase analysis in Ref.~\citen{Byst}.

In our previous works (see Refs.~\citen{Kup1,Kup2,Ber}), along with the usual variant of MDST, we considered its more sophisticated formulation with taking account of the effects of two-nucleon correlations through including the most essential intermediate excitations (IE) of the target nuclei as well as of the effects of the noneikonal corrections (NE) to the $p$--$A$ amplitude. For this reason, in Fig.~\ref{F1} we show a comparison not only with the curves calculated by the usual MDST, which is analogous to the present variant of the MDES model, but also with the results obtained by the improved MDST calculation allowing for IE and NE. As can be seen from Fig.~\ref{F1}, the calculations based on the MDES model and in the framework of the MDST without taking into account IE and NE yield not so much differing results at the considered incident proton energies. The most significant distinctions are observed for the scattering on $^{40}$Ca in the region of diffraction minima.

For the analyzing power $A_y \left( \theta  \right)$, there are certain distinctions of the results obtained in these two approaches, in the case of the scattering on $^{40}$Ca the MDES model providing somewhat more encouraging results, however for the scattering on $^{208}$Pb this curve rises above the experimental points with the scattering angle increase, as it also does in the case of MDST. Certain distinctions between these two models are also observed in the calculations of the spin rotation functions $Q\left( \theta  \right)$. On the whole, we may say that the above-developed approach MDES in its present form without taking account of the IE and NE corrections does not yield a significant improvement in describing the analyzed experimental data in the considered energy region. In Fig.~\ref{F1} the shown results of calculations on the basis of MDST with the allowance for the mentioned corrections clarify their role for the data description. The contribution of IE of nuclei becomes essential with the scattering angle increase and the NE corrections manifest themselves in smoothing the diffraction minima of the observables, and in general, the allowance for the IE and NE corrections improves the description of experimental data. From this comparison we can make a conclusion about the necessity of further developing the MDES model, in particular, in order to take into account the effects of IE of target nuclei and NE corrections also in the approach basing on the Dirac equation. As other possible refinements of this model, we can also mention the allowance for the contributions to the $p$--$A$ amplitude coming from spin-spin terms in the $NN$-amplitude as well as refining the formulae by retaining the contributions of the lower components in the nucleon wave functions calculated in the RMF approximation.

\section{Conclusion}

By analogy with the Glauber multiple diffraction scattering theory (MDST), being a highly effective and popular approach to analyzing processes of the nuclear-particle scattering on atomic nuclei at intermediate energies, in the present work an attempt has been made to provide a consistent relativistic description of the process of multiple scattering of the incident proton on nucleons of the target nucleus. For this purpose, a new model has been built, which is developed basing on the eikonal approximation for the Dirac equation and, therefore, can be called as the model of multiple Dirac eikonal scattering (MDES). New expressions have been obtained for the amplitudes of the elastic $p$--$A$ scattering on a spinless target nucleus basing on the consideration of the multiple scattering Watson series with making use of the eikonal approximation for the Dirac propagators of free proton movement between the acts of successive scattering on nucleons. In the framework of the developed MDES model an analysis of the complete set of the observables for the elastic $p+^{40}$Ca and $p+^{208}$Pb scattering has been performed at the incident proton energy of 800 MeV. This analysis is based on using the realistic nucleon densities, calculated by means of modern models for describing the nucleus structure in the approximation of relativistic mean field, and on using realistic $NN$-amplitudes obtained in the literature from the phase analysis.

A comparison has been made for the results of these calculations with the analogous calculations on the basis of both the usual variant of MDST and the more sophisticated variant of MDST with the allowance for intermediate excitations (IE) of the target nuclei and for noneikonal (NE) corrections. This comparison reveals certain slight distinctions between the results for MDES model and the usual variant of MDST and suggests that a further development of the MDES approach is advisable, in order to take account of the IE and NE corrections as well as to abandon several simplifications made during deriving the expressions for the scattering amplitude, in particular, to include contributions of the omitted spin-spin terms in the elementary $pN$-amplitude.

\end{document}